# Development of Beams Optics Code LEADS-v5[*]

Lü Jian-Qin[**]

State Key Laboratory of Nuclear Physics and Technology, Peking University, Beijing 100871, China

**Abstract**

To calculate nonlinear transport of space charge dominated beam in 6D phase spaces, a computer code package LEADS-v5 (Linear and Electrostatic Accelerator Dynamics Simulations) has been developed. The codes calculate particle motions in the beam transport systems consisting of electrostatic and magnetic focusing lenses, ion analyzers, multipoles and RF accelerating structures. The nonlinear forces of external electric/magnetic fields are analyzed by the Lie algebraic method, and the space charge forces are obtained by the particle in cell (PIC) scheme. In the codes, Uniform and Gaussian particle distributions can be chosen to generate randomly the particle initial coordinates. The optimization procedures are provided to make the beam optics designs reasonable and fast. Graphically displays of calculated results are provided.

## 1. Introduction

Intense beam nonlinear transport is a very complicated problem, because the state of particle motion is dominated not only by the applied electromagnetic fields, but also by the beam induced electromagnetic fields (self fields). Moreover, the self fields are related to the beam dimensions and particle distributions. So, it is impossible to get the self-consistent solutions of particle motions analytically. To solve this problem, we use two methods: Lie algebraic method and PIC method.

In 1987 Dr. Dragt first introduced the Lie algebraic theory [1], [2] into accelerator physics. Since then the theory has been widely used by accelerator physicists in the research of charged particle nonlinear motions. The PIC method was originally developed for fluid dynamics simulations [3, 4]; however, its greatest application is in plasma physics [5, 6]. The method began to be used for particle accelerator simulations two decades ago. We combine the Lie algebraic method and the PIC method together to simulate intense beam nonlinear transport. With the Lie algebraic method we analyze the particle nonlinear trajectories in the applied electromagnetic fields, and with the PIC algorithm we calculate the effects of space charge forces of the beam. Based on the two methods, we have developed a computer program LEADS-5.

Optimization plays an important role in the computational design for the particle accelerators and beam optical lines. It will not only save much computational time, but also make the design more reasonable. Many accelerator codes can perform optimization calculations. For example, program TRANSPORT by K. L. Brown [7] has powerful optimization ability. For this reason, we put some optimization subroutines in LEADS-5v, by using the Powell nonlinear optimization method.

The calculated results, such as the beam phase areas both in transverse and longitudinal directions as well as the beam envelopes can be displayed graphically on the computer monitor, so that it makes the calculated results simple and visual.

[*] Work supported by the National Natural Science Foundation of China, Contract No. 1057009, and by SRFDP (20070001001)

[**] Email: jqlu@pku.edu.cn

## 2. Particle distributions

The particle distributions can be selected by the user. Two kinds of distributions are provided: uniform and Gaussian distributions in the 6D (x, x', y, y', ΔE, Δϕ) phase spaces as the initial particle coordinates which are generated randomly.

## 3. Beam focusing, analyzing and accelerating elements

The following focusing, analyzing and accelerating elements are provided by LEADS-5:

### 3.1. Electromagnetic focusing elements, multipoles (to fifth order approximation)

- Electrostatic quadrupole.
- Solenoid magnets.
- Magnetic quadrupole.
- Sextupole magnets.
- Octupole magnets.
- Decapole magnets.
- Dodecapole magnets.

### 3.2. Ion analyzing elements (to fifth order approximation)

- Dipole magnets (including fringe fields).
- Cylindrical electrostatic analyzers.
- Spherical electrostatic analyzers.
- E×B analyzers.

### 3.3. Axisymmetric electrostatic lenses (to third order approximation) [9-12]

- Two - cylinder accelerating gap lenses.
- Three - cylinder Einzel lenses.
- Three-aperture Einzel lenses.
- DC accelerator columns.

### 3.4. RF accelerating cavities (to first order approximation)

- Quarter wave resonators (QWRs).
- Split loop resonators (SLRs).

### 3.5. Other elements

- Element rotation; Every physical element can be rotated about the z-axis in arbitrary angle (positive/negative).
- Arbitrary matrix. If you have known the transfer matrix of an optical element, you can put in the matrix to the code input data directly.
- Phase space plotting. The uses can plot the phase space diagrams of the beam in x-, y-, z- directions any where along the beam line.
- Beam envelope plotting. Beam envelope plotting is provided in the code.

## 4. Analysis calculation tools and formulas

### 4.1. Lie algebraic tools

In the canonical phase space $\xi = (x, p_x, y, p_y, z, p_z)$, if one doesn't take the space charge effects into count, the Hamiltonian of the particle motion is

$$H_t = -\left[m_0^2 c^4 + c^2(p_x - qA_x)^2 + c^2(p_y - qA_y)^2 + c^2(p_z - qA_z)^2\right]^{\frac{1}{2}} + q\psi \tag{1}$$

where $x$, $y$ and $z$ are the particle coordinates in the real space; $p_x$, $p_y$ and $p_z$ are the particle canonical momentum components; $A_x$, $A_y$ and $A_z$ are the magnetic vector components; $\psi$ is the electric potential; $q$ is the particle charge, $m_0$ is the particle rest mass. Now, take $\zeta = (x, x', y, y', \tau, p_\tau)$ as a new is canonical phase space, where $\tau = t - z/v_0$ is the time difference between the arbitrary particle and the reference particle, $p_\tau = p_t - p_t^0$, $p_t = -H_t$, $p_t^0$ is the value of $p_t$ for the reference particle. From phase space $\xi$ to phase space $\zeta$, transformation $\xi \rightarrow \zeta$ is canonical. In the new phase space $\zeta$, the Hamiltonian is

$$H = -\left[-(p_x - qA_x)^2 - (p_y - qA_y)^2 - (p_t + p_t^0 + q\psi)^2/c^2 - m_0^2 c^2\right]^{1/2} + qA_z - \frac{p_t + p_t^0}{v_0} \quad (2)$$

and in the phase space $\zeta = (x, x', y, y', \tau, p_\tau)$ the solutions of particle trajectories can be expressed as

$$\zeta_{f_i} = \sum_{j=1}^{6} M_{i,j} \zeta_j + \sum_{j=1}^{6}\sum_{k=1}^{6} S_{i,j,k} \zeta_j \zeta_k + \sum_{j=1}^{6}\sum_{k=1}^{6}\sum_{l=1}^{6} T_{i,j,k,l} \zeta_j \zeta_k \zeta_l +$$

$$\sum_{j=1}^{6}\sum_{k=1}^{6}\sum_{l=1}^{6}\sum_{m=1}^{6} U_{j,k,l,m} \zeta_j \zeta_k \zeta_l \zeta_m + \sum_{j=1}^{6}\sum_{k=1}^{6}\sum_{l=1}^{6}\sum_{m=1}^{6}\sum_{n=1}^{6} W_{j,k,l,m,n} \zeta_j \zeta_k \zeta_l \zeta_m \zeta_n + \ldots + \Delta \zeta_i,$$

$$i=1,\ldots,6 \quad (3)$$

where $\zeta_f$ is the particle final coordinate, $\zeta$ the initial coordinate; $M_{i,j}$, $S_{i,j,k}$, $T_{i,j,k,l}$, $U_{i,j,k,l,m}$ and $W_{i,j,k,l,m,n}$ are the first order, second order and to the fifth order coefficients of the particle trajectories contributed by the applied electromagnetic fields, $\Delta \zeta$ is the contributions of the space charge forces to the particle trajectories.

As mentioned before, the particle trajectories in the external fields are analyzed with the Lie algebraic method, and the effects of space charge forces are calculated with the PIC scheme.

Expanding the Hamiltonian $H$ into power series gives

$$H = H_0 + H_1 + H_2 + H_3 + H_4 + H_5 + H_6 + \ldots \quad (4)$$

In the phase space, the final coordinate $\zeta_f$ and the initial coordinate $\zeta$ of a particle are related by a map $M$:

$$\zeta_f = M \zeta \quad (5)$$

According to the references [1] and [2], M is

$$M = \cdots M_4 M_3 M_2 \quad (6)$$

where

$$M_2 = \exp(:f_2:), \quad M_3 = \exp(:f_3:), \quad M_4 = \exp(:f_4:), \quad M_5 = \exp(:f_5:), \quad M_6 = \exp(:f_6:), \cdots \quad (7)$$

and

$$f_2 = -\int_0^\ell H_2(z) \, dz \quad (8)$$

$$f_3 = -\int_0^\ell h_3^{\text{int}}(z) \, dz \quad (9)$$

$$f_4 = -\int_0^\ell h_4^{\text{int}}(z) \, dz \quad (10)$$

$$f_5 = -\int_0^\ell h_5^{\text{int}}(z_1)\,dz_1 + \int_{z_0}^\ell dz_1 \int_{z_0}^{z_1} dz_2 [-h_3^{\text{int}}(z_2),-h_4^{\text{int}}(z_1)]$$

$$+\frac{1}{3}\int_{z_0}^\ell dz_1 \int_{z_0}^{z_1} dz_2 \int_{z_0}^{z_2} dz_3 \big([-h_3^{\text{int}}(z_3),[-h_3^{\text{int}}(z_2),-h_3^{\text{int}}(z_1)]]$$

$$+[-h_3^{\text{int}}(z_2),[-h_3^{\text{int}}(z_3),-h_3^{\text{int}}(z_1)]]\big)$$ (11)

$$f_6 = -\int_0^\ell h_6^{\text{int}}(z_1)\,dz_1 + \frac{1}{4}\int_{z_0}^\ell dz_1 \int_{z_0}^{z_1} dz_2 [-h_3^{\text{int}}(z_2),-h_5^{\text{int}}(z_1)] + \frac{1}{2}\int_{z_0}^\ell dz_1 \int_{z_0}^{z_1} dz_2 [-h_4^{\text{int}}(z_2),-h_4^{\text{int}}(z_1)]+$$

$$\frac{1}{4}\bigg(\int_{z_0}^\ell dz_1 \int_{z_0}^{z_1} dz_2 \int_{z_0}^{z_2} dz_3 \big([-h_4^{\text{int}}(z_3),[-h_3^{\text{int}}(z_2),-h_3^{\text{int}}(z_1)]]$$

$$\int_{z_0}^\ell dz_1 \int_{z_0}^{z_1} dz_2 \int_{z_0}^{z_2} dz_3 \big([-h_4^{\text{int}}(z_2),[-h_3^{\text{int}}(z_3),-h_3^{\text{int}}(z_1)]] +$$

$$3\int_{z_0}^\ell dz_1 \int_{z_0}^{z_1} dz_2 \int_{z_0}^{z_2} dz_3 \big([-h_3^{\text{int}}(z_3),[-h_3^{\text{int}}(z_2),-h_4^{\text{int}}(z_1)]] +$$

$$\int_{z_0}^\ell dz_1 \int_{z_0}^{z_1} dz_2 \int_{z_0}^{z_2} dz_3 \big([-h_3^{\text{int}}(z_2),[-h_3^{\text{int}}(z_3),-h_4^{\text{int}}(z_1)]]\bigg)$$

$$\frac{1}{4}\bigg(\int_{z_0}^\ell dz_1 \int_{z_0}^{z_1} dz_2 \int_{z_0}^{z_2} dz_3 \int_{z_0}^{z_3} \big([-h_3^{\text{int}}(z_3),[-h_3^{\text{int}}(z_2),[-h_3^{\text{int}}(z_4),-h_3^{\text{int}}(z_1)]]] +$$

$$\int_{z_0}^\ell dz_1 \int_{z_0}^{z_1} dz_2 \int_{z_0}^{z_2} dz_3 \int_{z_0}^{z_3} \big([-h_3^{\text{int}}(z_4),[-h_3^{\text{int}}(z_2),[-h_3^{\text{int}}(z_3),-h_3^{\text{int}}(z_1)]]] +$$ (12)

$$\int_{z_0}^\ell dz_1 \int_{z_0}^{z_1} dz_2 \int_{z_0}^{z_2} dz_3 \int_{z_0}^{z_3} \big([-h_3^{\text{int}}(z_4),[-h_3^{\text{int}}(z_3),[-h_3^{\text{int}}(z_2),-h_3^{\text{int}}(z_1)]]]\bigg)$$

where $\ell$ is the length of the element. and

$$h_n^{\text{int}}(\varsigma,z) \quad M_2 H_n \tag{13}$$

To simplify the expressions we define the multiple integrations as follows:

$$\left\langle \begin{matrix} i & j & m \\ \alpha, & \beta, & \cdots, & \varepsilon \end{matrix} \right\rangle = \int_{t^{in}}^t dt_i \int_{t^{in}}^{t_i} dt_j \cdots \int_{t^{in}}^{t_l} dt_m \left[ -H_\alpha^{\text{int}}(t_i) :: -H_\beta^{\text{int}}(t_j) : \cdots : -H_\varepsilon^{\text{int}}(t_m) : \right] \tag{14}$$

Then, the equations (9) — (12) can be rewritten as

$$f_3 = \begin{bmatrix} 1 \\ 3 \end{bmatrix} \tag{15}$$

$$f_4 = \begin{bmatrix} 1 \\ 4 \end{bmatrix} + \frac{1}{2}\begin{bmatrix} 2 & 1 \\ 3, & 3 \end{bmatrix} \tag{16}$$

$$f_5 = \begin{bmatrix} 1 \\ 5 \end{bmatrix} + \begin{bmatrix} 2 & 1 \\ 3, & 4 \end{bmatrix} + \frac{1}{3}\begin{bmatrix} 3 \\ 3, & \begin{bmatrix} 2 & 1 \\ 3, & 3 \end{bmatrix} \end{bmatrix} + \frac{1}{3}\begin{bmatrix} 2 \\ 3, & \begin{bmatrix} 3 & 1 \\ 3, & 3 \end{bmatrix} \end{bmatrix} \tag{17}$$

$$f_6 = \begin{bmatrix} 1 \\ 6 \end{bmatrix} + \begin{bmatrix} 2 & 1 \\ 3, & 5 \end{bmatrix} + \frac{1}{2}\begin{bmatrix} 2 & 1 \\ 4, & 4 \end{bmatrix} + \frac{1}{4}\left(\begin{bmatrix} 3 \\ 4, & \begin{bmatrix} 2 & 1 \\ 3, & 3 \end{bmatrix} \end{bmatrix} + \begin{bmatrix} 2 \\ 4, & \begin{bmatrix} 3 & 1 \\ 3, & 3 \end{bmatrix} \end{bmatrix} + 3\begin{bmatrix} 3 \\ 3, & \begin{bmatrix} 2 & 1 \\ 3, & 4 \end{bmatrix} \end{bmatrix} + \begin{bmatrix} 2 \\ 3, & \begin{bmatrix} 3 & 1 \\ 3, & 3 \end{bmatrix} \end{bmatrix}\right)$$

$$+\frac{1}{4}\left(\begin{bmatrix} 3 \\ 3, & \begin{bmatrix} 2 \\ 3, & \begin{bmatrix} 4 & 1 \\ 3, & 3 \end{bmatrix} \end{bmatrix} \end{bmatrix} + \begin{bmatrix} 4 \\ 3, & \begin{bmatrix} 2 \\ 3, & \begin{bmatrix} 3 & 1 \\ 3, & 3 \end{bmatrix} \end{bmatrix} \end{bmatrix} + \begin{bmatrix} 4 \\ 3, & \begin{bmatrix} 3 \\ 3, & \begin{bmatrix} 2 & 1 \\ 3, & 3 \end{bmatrix} \end{bmatrix} \end{bmatrix}\right) \tag{18}$$

The first order, second order, and to fifth order terms of the orbit solutions ($\zeta^i$, i=1,2,…,5,…) are

expressed as

$$\varsigma^1 = f_2\varsigma \tag{19}$$

$$\varsigma^2 = f_3\varsigma^1 \ldots\ldots \tag{20}$$

$$\varsigma^3 = (:f_4: + \frac{1}{2}:f_3:^2)\varsigma^1 \tag{21}$$

$$\varsigma^4 = (:f_5: + :f_4::f_3: + \frac{1}{6}:f_3:^3)\varsigma^1 \tag{22}$$

$$\varsigma^5 = (:f_6: + :f_5::f_3: + \frac{1}{2}:f_4:^2 + \frac{1}{2}:f_4::f_3:^2 + \frac{1}{24}:f_3:^4)\varsigma^1 \tag{23}$$

where :*: stands for the Poisson operation. For example, for the given two functions $f$ an $g$, $:f: g = [f, g]$.

## 4.2. The PIC algorithms [5, 6]

The beam self-fields are calculated with the PIC scheme. The initial particle distributions are generated randomly.

Generally speaking, to simulate the particle motion in the beam self-field with the PIC method, the following steps would be taken:

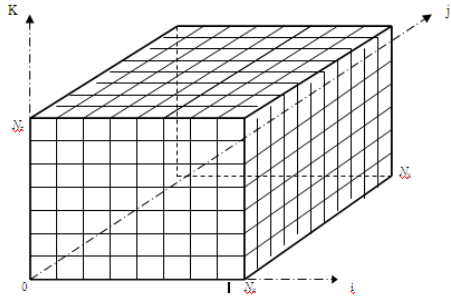

**Fig. 1 Mesh generation**

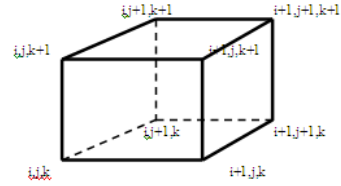

**Fig 2 a mesh cell**

a. Randomly generate the particle initial distributions.

b. Suppose the beam moves along the straight line (ignoring the curvilinear orbit due to the short time step $\Delta t$); Divide the beam into cubic mesh cell (see Figs. 1 and 2); Let $h_x$, $h_y$, and $h_z$ be the grid widths in the x, y and z directions respectively; the grid node numbers are $N_x$, $N_y$ and $N_z$ in the x, y and z directions respectively.

c. Assign the charge $q$ of each macroparticle (MP) in the cell to the cell nodes: the charge of every MP is distributed to the eight cell nodes. The portion of $q$ assigned to the cell nodes is determined by the position of the MP within the cell in accordance with the following relationship:

$$\rho_{l,m,n} = \sum_{p=1}^{N_c} \rho_{MP}\left(1 - \frac{|x_p - x_l|}{h_x}\right)\left(1 - \frac{|y_p - y_m|}{h_y}\right)\left(1 - \frac{|z_p - z_n|}{h_z}\right) \tag{24}$$

where $l = i, i+1$, $m = j, j+1$, $n = k, k+1$ are integer grid indices; $N_c$ is MP number in this cell; $p$ is the sequence number of MP in the cell; $(x_p, y_p, z_p)$ are the coordinates in the cell, $(x_l, y_m, z_n)$ are the coordinates of a node; and

$$\rho_{xyz} = \frac{q}{h_x h_y h_z}. \tag{25}$$

From Eq. (a) we see that the MP is not a point, but has a finite volume $h_x \times h_y \times h_z$, and $\rho_{MP}$ is just the charge density of a MP.

d. Solve the Poisson equation $\nabla^2 \phi = -\rho(x,y,z)/\varepsilon_0$ ($\varepsilon_0$ is the vacuum permittivity): After the charge of all of the MPs has been assigned to the mesh nodes, we use the Green's function to calculate the potentials $\phi_{i,j,k}$ at the mesh nodes. Usually, the beam transverse dimensions are much smaller than the vacuum pipe, so we can use the opening boundary conditions, and take $\phi_{r \to \infty} = 0$.

The solutions of the Poisson equation are expressed as:

$$\phi(x,y,z) = -\frac{1}{4\pi\varepsilon_0} \int G(x,x_0,y,y_0,z,z_0)\rho(x_0,y_0,z_0)dx_0 dy_0 dz_0 \tag{26}$$

here $(x,y,z)$ is field point, $(x_0, y_0, z_0)$ is source point, and the Green's function is

$$G(x,x_0,y,y_0,z,z_0) = \frac{1}{\sqrt{(x-x_0)^2 + (y-y_0)^2 + (z-z_0)^2}} \tag{27}$$

Rewrite Eq. (27) to discrete form, we get the potentials at the mesh nodes:

$$\phi(x_i, y_j, z_k) = -\frac{h_x h_y h_z}{4\pi\varepsilon_0} \sum_{i'=1}^{N_x} \sum_{j'=1}^{N_y} \sum_{k'=1}^{N_z} G(x_i - x_{i'}, y_j - y_{j'}, z_k - z_{k'}) \rho(x_{i'}, y_{j'}, z_{k'}) \tag{28}$$

e. The electric fields at the mesh nodes are calculated with the central interpolation method

$$E_x = -\frac{\phi_{i+1,j,k} - \phi_{i-1,j,k}}{2h_x} \quad E_y = -\frac{\phi_{i,j+1,k} - \phi_{i,j-1,k}}{2h_y} \quad E_z = -\frac{\phi_{i,j,k+1} - \phi_{i,j,k-1}}{2h_z}. \tag{29}$$

The electric fields at all MP positions are calculated by interpolating from the electric field at the mesh nodes using the same weighting scheme as what for the charge assignment.

f. From the Newton's law, the contributions of the self-fields to the particle trajectories are:

$$\Delta x = \frac{1}{2} q E_x \Delta t^2 / m, \Delta y = \frac{1}{2} q E_y \Delta t^2 / m, \Delta z = \frac{1}{2} q E_z \Delta t^2 / m. \tag{30}$$

where $m$ is the particle mass, $\Delta t$ $\Delta z/v_0$, $\Delta z$ is the step length along the reference orbit, $v_0$ is the velocity of the reference particle. The particle velocity changes are

$$\Delta v_x = q E_x \Delta t / m, \Delta v_y = q E_y \Delta t / m, \Delta v_z = q E_z \Delta t / m. \tag{31}$$

Combining the results obtained both from the PIC scheme and the Lie algebraic method, we get the particle trajectories in the optical elements.

### 4.3. RF gap formulas
### 4.3.1 Single RF gap

The code provides the beam transport calculations for the RF accelerating structures consisting of QWR or SLR cavities, see Figs. 3 and 4.

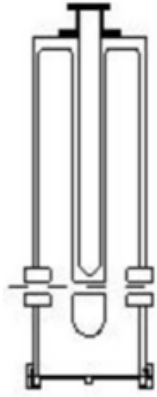
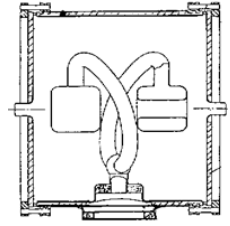

              Fig. 3 Quarter wave resonator        Fig. 4 Split loop resonator

For each RF gap of QWR or SLR the transfer matrix is

$$M = M(i, j) \quad\quad i, j = 1,6 \quad\quad (32)$$

where

$$M_{11}=1-(\alpha/2)(T_k k/\gamma^3+T/\gamma)\cos\varphi$$
$$M_{12}=-(\alpha/2-T_{kk}k/\gamma^3)\sin\varphi$$
$$M_{21}=-(\alpha k T/(2\gamma^3))\sin\varphi$$
$$M_{22}=1-(\alpha/\gamma)(1/\gamma^2+\beta^2)\,T\cos\varphi$$
$$M_{33}=M_{11} \quad M_{34}=M_{12} \quad\quad (33)$$
$$M_{43}=M_{21} \quad M_{44}=M_{22}$$
$$M_{55}=1+(\alpha\,k\,T_k/\gamma^3)\cos\varphi$$
$$M_{65}=-QTV\sin\varphi$$
$$M_{66}=1$$

The matrix entries not listed above are all zero, and Q is the number of particle charge state, $V_m$ is the voltage amplitude across the RF gap, $\alpha=QV/(2E)$, $E$ the energy of reference particle, $k=\omega/v$, $\omega$ is the frequency of the RF field, v is the velocity of the reference particle, T, $T_k$ and $T_{kk}$ are the time transit factor and its first and second deviations to k, $\gamma$ is the relativistic factor.

### 4.3.2 Quarter wave resonators

To calculate the transfer matrix at each gap the energy of reference particle should known first. For the QWRs, let $\phi_s$ is the phase angle of the reference particle at the middle way of a QWR, the energy gain of reference particle at the first gap is:

$$\Delta E = Q \cdot V_m \cdot T_{tf} \cdot \cos(\phi_s + \pi/2 - \omega\ell/v + \phi_1) \quad\quad (34)$$

where $T_{tf}=[\sin(\omega g/2v)/(\omega g/2v)]$ is the time transit factor; $2\ell$ is the length between the two centers of the two gaps; $\phi_1$ is the initial phase angle of the reference particle when arriving at the first gap. In the above equation the particle energy depends on its phase angle, and the phase angle is related to the energy (in other word, the velocity v). So, (34) is a transcendental equation and can be solved iteratively. From the gap 1 to gap 2 the particle drift distance $2\ell$, the energy keeps unchanged, but the phase angle increases by $2\ell\omega/v$. The energy gain of reference particle in gap 2 is

$$\Delta E = Q \cdot V_m \cdot T_{tf} \cdot \cos(\varphi_s - \pi/2 - \omega\ell/v + \varphi_2) \quad\quad (35)$$

where $\varphi_2$ is the phase angle of reference particle at gap 2. At this time the velocity $v$ is a constant, it is not necessary to calculate the energy gain iteratively.

### 4.3.3 Split loop resonator

For the SLR element, at the first gap the phase angle and energy gain of the reference particle can be calculated iteratively according to Eq. (34). At the second gap the phase angle is $\phi_s$, the energy gain is

$$\Delta E = Q \cdot V_m \cdot T_{tf} \cdot \cos\varphi_s \tag{36}$$

When the particle arrives at the third gap, its phase angle and energy gain can be obtained from Eq. (35) directly.

## 5. Optimization procedure [8]

We have developed the optimization subroutines which automatically determine the optimal parameters of the optical elements to match a beam from a given initial state to a prescribed final state.

Several optical conditions can be prescribed, such as forming an image, making a beam waist, chromaticity, etc. If the transfer matrix of an element or a part of the beam line is M (i, j), i, j=1, 6, the beam matrix is σ(i, j), i, j=1, 6, the following optical conditions could be inserted into the input data file:

- image in x-plane: M(1, 2)=0;
- image in y-plane: M(3, 4)=0;
- chromaticity: M(1, 6)=0, or M(2, 6)=0;
- Parallel—Focusing image: M(1, 1)=0;
- Focusing—Parallel image: M(2, 2)=0;
- Parallel—Parallel image (telescope system): M(2, 1)=0;
- form a waist in x-plane: σ(1, 2)=0;
- beam size limit in x-plane: σ(1, 1)=given value;
- form a waist in y-plane: σ(3, 4)=0;
- beam size limit in y-plane: σ(3, 3)= given value;
- beam waist in the longitudinal direction: σ(5, 6)=0.

Powell nonlinear optimization subroutines [4] have been incorporated in the codes. The goal of the optimization calculations is to find out the minimum values of the following object function

$$F = \sum_{i=1}^{n} \left[ (f_i(x_1, x_2, \cdots, x_m) - f_{i0})/\varepsilon \right]^2 = \min \tag{37}$$

where $f_i$ (i=1,2,...n) are the required optical conditions; fin are the given value for this conditions; $x_j$ (j=1,2,...,m) are the variable parameters, such as magnetic field, voltage, element length and so on; $\varepsilon_i$ is the tolerance for each condition (weight factor).

## 6. Periodical structure calculation [9]

In the linear particle accelerators consisting of QWRs or SLRs, or in some periodically arranged beam lines, the particle beams will pass through these periodic structures. In order to keep the particle motion stable, the program automatically adjusts the magnetic quadrupole fields to fit the following stability condition:

$$|\cos\mu| = |0.5 \operatorname{Tr}(\mathbf{M})| \leq 1, \tag{38}$$

where $\mu$ is the phase shift per period,     is the Twiss matrix shown as the follows:

$$\mathbf{M} = \begin{pmatrix} \cos\mu + \alpha\sin\mu & \beta\sin\mu \\ -\gamma\sin\mu & \cos\mu - \alpha\sin\mu \end{pmatrix}. \tag{39}$$

The periodic structures could be combined with the magnetic quadrupoles, drift spaces, as well as QWRs and SLRs.

## 7. Simulation examples

In this section we present four example applications of the code. The first example is a low energy beam transport system (LEBT) of the RFQ accelerator. By this example we just want to illustrate the space charge effects to the beam transport. The second one is the optical system of the 400 keV high voltage accelerator. Most of elements of the accelerator are Axisymmetric electrostatic lenses. The third one is a particle distribution uniformization system, which exhibits the applications of multipoles. The last one is an RF linear accelerator system combined with QWRs and quadrupoles.

### 7.1 The LEBT system of a RFQ accelerator

With the code, we calculated the LEBT system which delivers the $D^+$ beam to the RFQ accelerating cavity (see Fig.5).

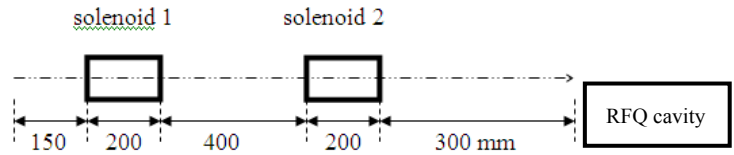

**Fig.5 Layout of the LEBT**

The LEBT system consists of two solenoid lenses and some drift spaces. The radius the solenoids are 50 mm, the lengths of them are all 200 mm, the longitudinal magnetic fields in the two solenoids are all 4.77 kG.

The $D^+$ beam energy is 50 keV, the average bunched beam current is 5 mA, the initial parameters of the beam in the phase space are 1×40 mm−mrad both in the $x$− and $y$−directions. The pulse repetition is 9 MHz, and the initial longitudinal phase space ($\delta E \times \delta\phi$) is 1 keV×60°. The calculated beam envelopes of the system are shown in Fig. 6.

We can see from Fig. 6 that the beam envelopes of 0 mA beam current are much different from the envelopes of 5 mA beam current.

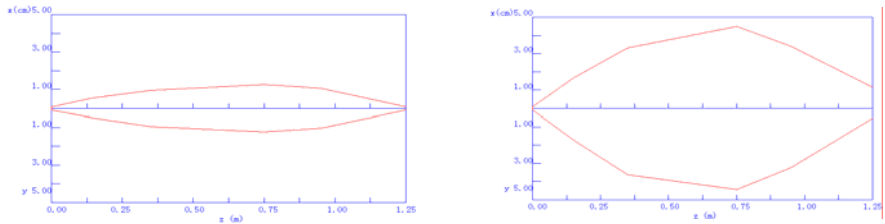

(a) Beam envelopes with 0 mA current;   (b) Beam envelopes with 5 mA current.
**Fig.6 Beam envelopes in the LEBT system**

## 7.2 The 400 keV high voltage accelerator

The second example is an optical system of the 400 keV high voltage accelerator (Fig. 7). The initial proton beam parameters are: x=y=± 1 mm, x′=y′=± 40 mrad, δE=± 150 eV, δϕ=± 60°. The Einzel lens 1 focuses the beam to the center of the gap lens 2 which matches the beam to the high voltage accelerating columns. Usually, the beam emerging from the accelerating columns is divergent, so a quadrupole doublet 4 focuses the beam again to the target. Fig. 8 shows the beam envelopes in the optical system.

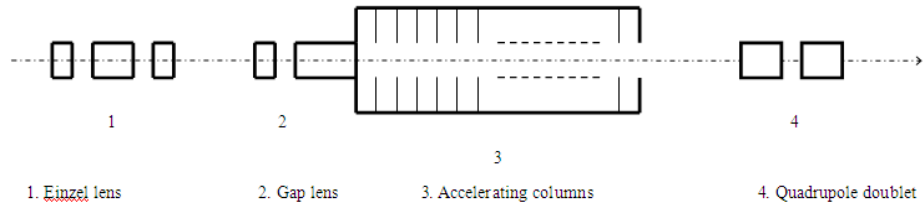

**Fig.7 Layout of the 400 keV high voltage Accelerator**

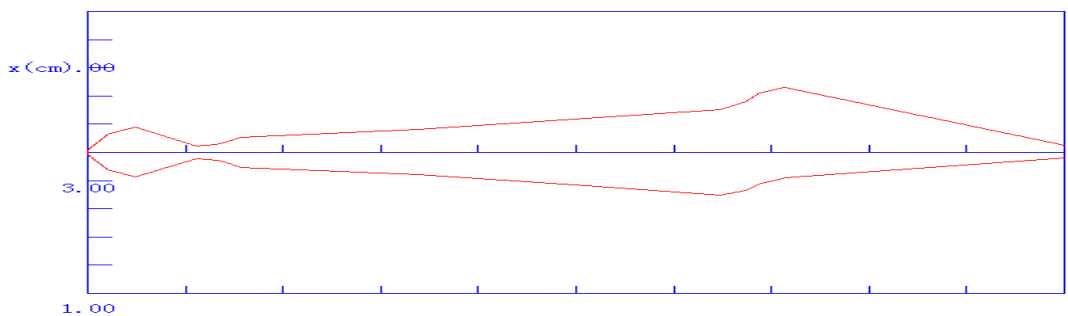

**Fig. 8 Beam envelopes in the 400 keV high voltage Accelerator**

## 7.3 Particle uniform distribution system

The third example of the applications of the code is a particle uniform distribution system. It is made up of two quadrupole triplets, one quadrupole doublet and two octupoles as shown in Fig. 9. The first octupole locates at a beam waist in x - plane; another octupole locates at the beam peak.

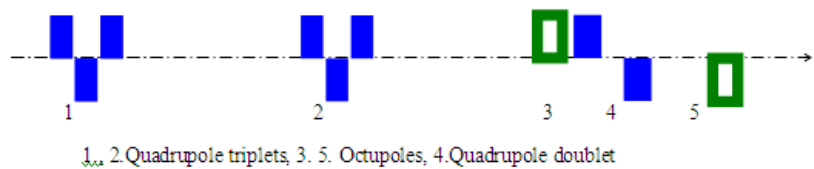

**Fig.9 Particle distribution uniformization system**

The initial proton beam energy is 400 KeV. The initial particle distribution in the x-y transverse plane is Gaussian. The initial beam parameters in the 6D phase space are (x, x′)=(±1 mm, ±20 mrad), (y, y′)=(±1 mm, ±20 mrad), (δE, δϕ) = (±150 eV,±60°). The beam envelopes are shown in Fig. 10, the initial and final transverse distributions are shown in Fig.11.

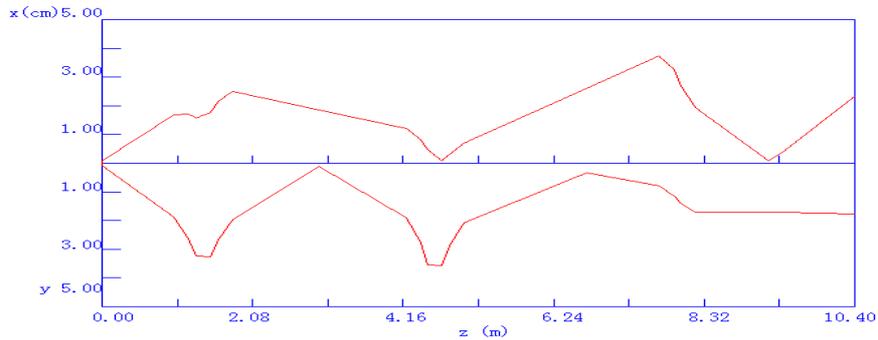

**Fig.10 Beam envelopes in the uniform distribution system**

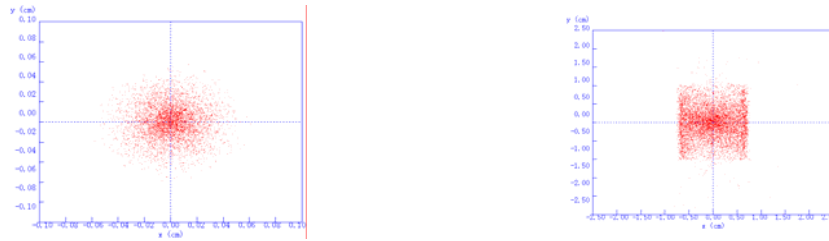

**(a) Initial, Gaussian distribution.   (b) Final, approximately uniform and square.**

**Fig.11 Particle distributions in transverse plane (x, y)**

### 7.4 The RF linear accelerator

The fourth example is a proton RF linear accelerator consisting of some QWRs. The accelerator consists of an array of 24 QWRs distributed in 6 cryostat modules (Fig.12). Each module contains 4 QWRs of the same structure and same physical parameters. The designed β of the linac is 0.1945, synchrotron phase angle –26.0°. The accelerating electric field of each QWR is 3.0 MV/m, working frequency 150 MHz. Fig.13 is the layout of the linac.

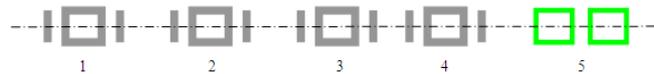

**Fig.12 Structure of a module**

1, 2, 3, 4: QWRs, 5: Quadrupole doublet

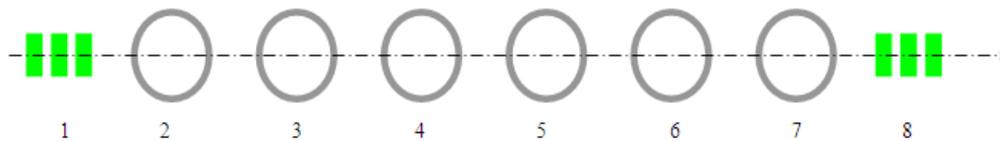

**Fig.13 Layout of the linac**

1, 8: Quadrupole triplets; 2, 3, 4, 5, 6, 7: Modules

The initial proton beam energy is 18.13 MeV. The initial beam parameters in the 6D phase space are (x, x′)=(±1 mm, ±5 mrad), (y, y′)=(±1 mm, ±5 mrad), (δE, δϕ) = (±300 eV,±60°). The beam envelopes in the linac are shown in Fig. 14.

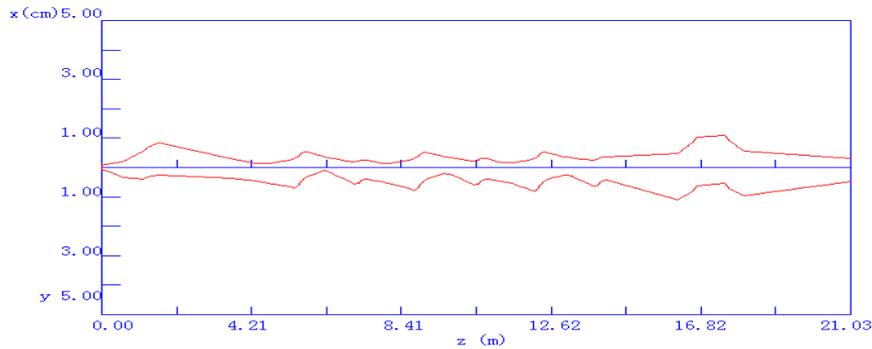

**Fig.14 Beam envelopes in the QWR linac**

**8. Conclusions**

A code package LEADS- v5 for nonlinear beam transport study with space charge effects has been developed. The code calculates high voltage accelerators which consist of axisymmetric electrostatic lenses and accelerating columns. Beam transport lines and RF linacs of QWR/SLR structures can be simulated also. The code provides multipole magnets which allows the user study the particle distribution uniformization systems. The optimization procedures provide the possibility to obtain the prescribed optical condition automatically. Many calculated examples show that the code is reliable.

**References**
[1] Dragt A.J., Lectures on nonlinear orbit dynamics, **AIP Conference Proceedings** No.87, edited by Carrigan R.A. et al. (**Am. Inst. Phys.** ,New York), (1982): 147–269
[2] Dragt A.J., and J.M. Finn, Lie Series and Invariant Functions for Analytic Symplectic Maps, **J. Math Phys**, 1976, 17(12): 2215.
[3] Buneman O., Phys. Rev., **115**, 503-517 (1959).
[4] Dawson J.M., Phys. Fluids, **5**, 445-459 (1962).
[6] Birdsall C.K., Clouds-in-clouds, clouds-in-cells physics for many-body plasma simulation, **Journal of Computational Physics**, 135, 1997: 141-148.
[6] Hockney R.W. and Eastwood J.W., Computer Simulation Using Particles, **Adam Hilger**, **Bristol**, 1987.
[7] Brown K.L., Rothacker F., Carey D.C. and Iselin Ch., Report SLAC-91, 1974.
[8] Himmelblau D.M., 1972 **Applied Nonlinear Programming** (New York: MC Graw-Hill).
[9] Adams A. and Read F.H., J. Phys. E 5 (1972) 1.50.
[10] Grivet P., **Electron Optics** (Pergamon, New York, 1965).
[11] Harting E. and Read F.H., **Electrostatic Lenses** (Elsevier, Amsterdam, Oxford, New York, 1976).
[12] Courant E.D. and Snyder M.S., **Ann. Phys**. 3 (19.58) 1.